\documentclass[letter,scriptaddress,twocolumn, prl,showkeys,showpacs,superscriptaddress]{revtex4-1}

	\usepackage{amsmath}
	\usepackage{makeidx}
	\usepackage{amsfonts}
	\usepackage{mathrsfs}
	\usepackage{graphicx, xcolor}  
	\usepackage{dcolumn}   
	\usepackage{bm}        
	\usepackage{bbm}
	\usepackage{amssymb}   
	\usepackage[ansinew]{inputenc}
	\usepackage[usenames,dvipsnames]{pstricks}
	\usepackage{subfigure}
	\usepackage{epstopdf}
	\usepackage{pst-grad} 
	\usepackage{pst-plot} 
	\usepackage[colorlinks,hyperindex]{hyperref}
	\usepackage[normalem]{ulem}
	\hypersetup
	{
		colorlinks,%
		citecolor=black,%
		linkcolor=black,%
		urlcolor=black,%
	}

\newcommand{\be}{\begin{equation}}
\newcommand{\ee}{\end{equation}}
\newcommand{\ba}{\begin{array}}
\newcommand{\ea}{\end{array}}
\newcommand{\bea}{\begin{eqnarray}}
\newcommand{\eea}{\end{eqnarray}}
\newcommand{\bsp}{\begin{split}}
\newcommand{\esp}{\end{split}}
\newcommand{\bmm}{\begin{multline}}
\newcommand{\emm}{\end{multline}}

\newcommand{\rd}{{\rm d}}

\begin{document}

\title{Edge Mode Amplification in Disordered Elastic Networks}
\author{Le Yan\thanks{lyan@kitp.ucsb.edu}}
\affiliation{Kavli Institute for Theoretical Physics, University of California, Santa Barbara, CA 93106, USA}
\author{Jean-Philippe Bouchaud}
\affiliation{Capital Fund Management - 23, rue de l'Universite 75007 Paris - France}
\author{Matthieu Wyart}
\affiliation{Institute of Physics, EPFL, CH-1015 Lausanne, Switzerland}

\date{\today}

\begin{abstract}
We study theoretically and numerically the propagation of a displacement field imposed at the edge 
of a disordered elastic material. While some modes decay with some inverse penetration depth $\kappa$, 
other exponentially {\it amplify} with rate $|\kappa|$, where $\kappa$'s are Lyapounov exponents analogous to those governing electronic transport in a disordered conductors. We obtain an analytical approximation for the full distribution $g(\kappa)$, which decays exponentially for large $|\kappa|$ and is finite when  $\kappa\rightarrow0$. Our analysis shows that isostatic materials generically act as levers with possibly very large gains, suggesting a novel principle to design molecular machines that behave as elastic amplifiers.

\end{abstract}

\maketitle

\section{Introduction}

The elastic response in amorphous solids  reveal several mesmerizing phenomena at a microscopic scale, including the presence of force chains \cite{Liu95,Majmudar05,Cates98} and the abundance of soft elastic modes that affect  elastic response \cite{Wyart05b,Leonforte06,Lerner14}, control  the material's stability \cite{Wyart05a,DeGiuli14} and its energy landscape \cite{Charbonneau14,Franz15,Biroli16}. How disorder structures respond to a perturbation is also fundamental in biology: many proteins are ``allosteric'', \textcolor{black}{meaning that}
 binding of one ligand mechanically affects other far-away sites specifically, up to several tenths atomic sizes  \cite{Changeux05,Zheng06}. Recently, it was found that certain elastic networks display zero modes at their edges, whose amplitude can decay or grow exponentially as one penetrates into the bulk \cite{Moukarzel12,Kane14}. Such modes can lead to intriguing properties such as very large, essentially non-linear response \cite{Kane14,Chen14}. For some crystalline lattices, topological considerations explain this behavior \cite{Kane14,Lubensky15}. However, the phenomenon appears to be more general \cite{Sussman15,Witten16} and is found in isostatic random networks (i.e. for which the coordination $z$ is such that the number of degrees of freedom $Nd$, where $d$ is the spatial dimension and $N$ the number of nodes, equals  the number of constraints $N_c$, the number of bonds between nodes, leading to  $z=2d$ \cite{Maxwell64}). One well studied example  is provided by random packings of hard particles with an average number of contacts per particle $z=2d$ \cite{Liu10,OHern03}. Denoting by $\kappa$ the inverse amplification/decay length of an edge mode, it is empirically found that the distribution $g(\kappa)$ has a thermodynamic limit \cite{Sussman15}, which is currently unexplained. 

In this Letter we argue that this problem can be rephrased as the Singular Value Decomposition (SVD) of a product of a large number of ``transfer'' matrices, each describing the transmission of zero modes through a thin slab of an isostatic material. This approach is similar to that used to describe electronic transport in quasi-one dimensional systems (see e.g. \cite{Beenakker97}, section III). We first argue, and confirm numerically, that the SVD spectrum of a single slab is 
well described by a Cauchy distribution, with a power-law tail arising from to the presence of specific (``lever'') arrangements of nodes that can considerably amplify the displacements. We then compute the distribution $g(\kappa)$ analytically in the thermodynamic limit, under the approximation that the different slab matrices are mutually free (i.e. their eigendirections are independent). Our prediction for $g(\kappa)$ agrees  well with observations, except in the vicinity of $\kappa = 0$. We discuss the cause of this discrepancy and show that our approximation becomes accurate in this region as well if some particles are pinned. We discuss several interesting implications of our findings.

\begin{figure}[h!]
\centering
\includegraphics[width=1\columnwidth]{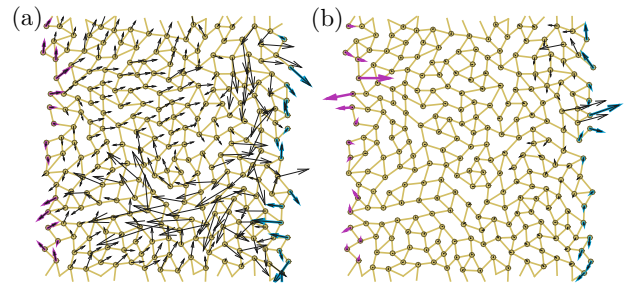}
\caption{ Response (\textcolor{black}{cyan arrows on the right boundary and black arrows in the bulk}) to an imposed displacement on the left boundary (\textcolor{black}{magenta arrows}) in an isostatic network (yellow bonds and nodes) with free boundary condition in the horizontal direction and periodic  boundary condition in the vertical one.  Here two specific eigenvectors  of the transmission  matrix ${\cal T}_{RL}$ are shown with singular value $\lambda=\textcolor{black}{2}$ (a) and $\lambda= \textcolor{black}{20000}$ (b). For (b), the magnitude of the response field is decreased by a  factor $\textcolor{black}{{20000}}$ for visibility. See the  S.I. for details.
\label{magnif}}
\end{figure}

{\it Set up:} We consider a disordered a $d$-dimensional isostatic network  with $z=2d$ in the bulk, and with two open boundaries along the horizontal direction, as illustrated in Fig.~\ref{magnif}. For simplicity, we choose periodic boundary conditions in the other directions.  We seek to characterize how an imposed displacement field $|\delta{\bf R}_L\rangle$ at the left open boundary dictates the displacement field $|\delta{\bf R}_R\rangle$ on the right side of the system. Examples of such responses are shown in Fig.~\ref{magnif}.

\begin{figure}[h!]
\centering
\includegraphics[width=.8\columnwidth]{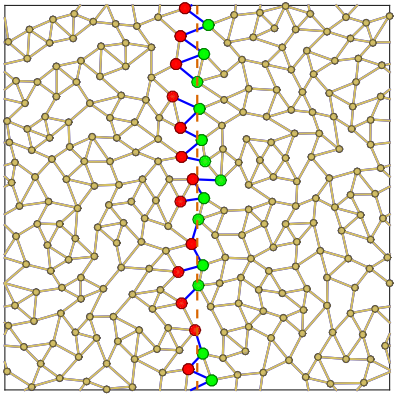}
\caption{An isostatic 2-dimensional network $z=4$ with periodic boundaries in both directions is cut at some $x_0$. Cut bonds are shown in blue, the boundary nodes are colored in green ($x>x_0$ left boundary) and red ($x<x_0$ right boundary) depending on the spatial positions relative to the cut. \label{cut}}
\end{figure}

To construct such a system one can consider a bulk isostatic material of linear size $L$ (in what follows, the average bond length defines our unit length), and cut all the contacts that cross a plane, as illustrated in Fig.~\ref{cut}.  This operation removes $\sim L^{d-1}$ constraints, and therefore generates as many floppy (zero-energy) modes. Note that some of these floppy modes will be strictly localized (with no displacement except near one of the open boundaries), but a finite fraction of them propagate throughout the system \cite{Wyart05a,Sussman15,Witten16}. We seek to characterize the displacements in the vector spaces of these extended floppy modes, propagating from the left (L) or from the right (R), of dimension $M = cL^{d-1}$, where $c<1$ is  constant that depends in general of the network considered.   Within this vector space, if the displacement of $M$ degrees of freedom are specified on (say) the left boundary, the floppy mode is entirely determined  throughout the bulk and projects onto the $M$ degrees of freedom on the right.

In the linear regime, the response $|\delta{\bf R}_R\rangle$ can therefore be written as:
\be
|\delta{\bf R}_R\rangle={\cal T}_{RL}|\delta{\bf R}_L\rangle
\label{eq_lin}
\ee
where  ${\cal T}_{RL}$ is a square $M \times M$ transmission matrix, once restricted to extended modes. How to extract ${\cal T}_{RL}$ numerically is described in S.I.. Transmission from right to left is obviously described by the matrix  ${\cal T}_{LR}={\cal T}_{RL}^{-1}$. Note that the transmission matrix ${\cal T}_{RL}$ is in general not symmetric and must be decomposed into its singular form:
\be
{\cal T}_{RL}=\sum_{\alpha=1...M} \Lambda_\alpha |\delta{\bf R_R}_\alpha\rangle \times \langle \delta{\bf R_L}_\alpha| 
\label{svd}
\ee
where the $\Lambda_\alpha$ are the \textcolor{black}{singular values~\footnote{The SVD is computed with Multiprecision Computing Toolbox~\cite{advanpix}.}}, $ |\delta{\bf R_R}_\alpha\rangle$ the left and  $| \delta{\bf R_L}_\alpha\rangle$ the right eigenvectors, both constituting an orthonormal basis. 

\begin{figure}[h!]
\centering
\includegraphics[width=.8\columnwidth]{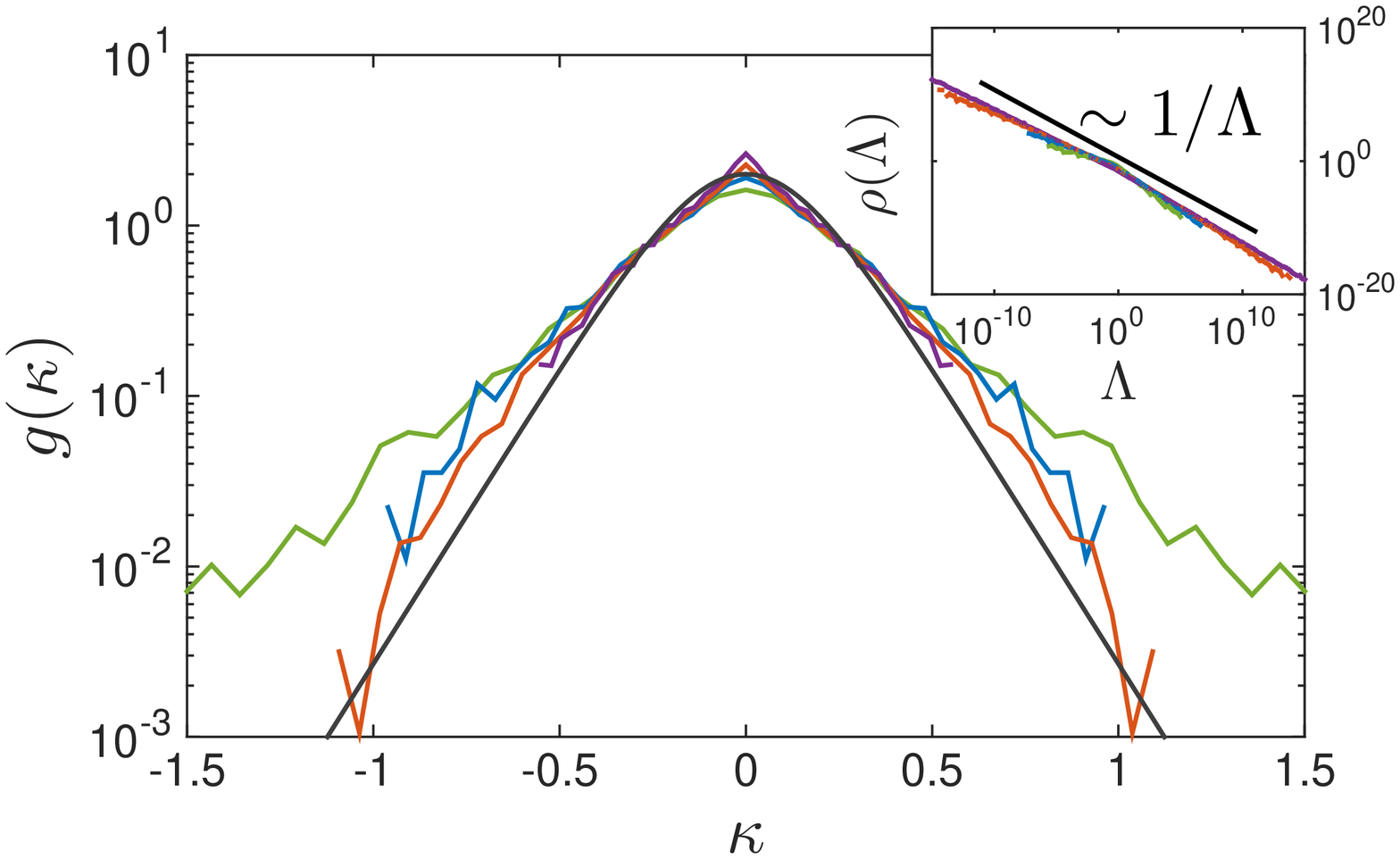}\\
\includegraphics[width=.8\columnwidth]{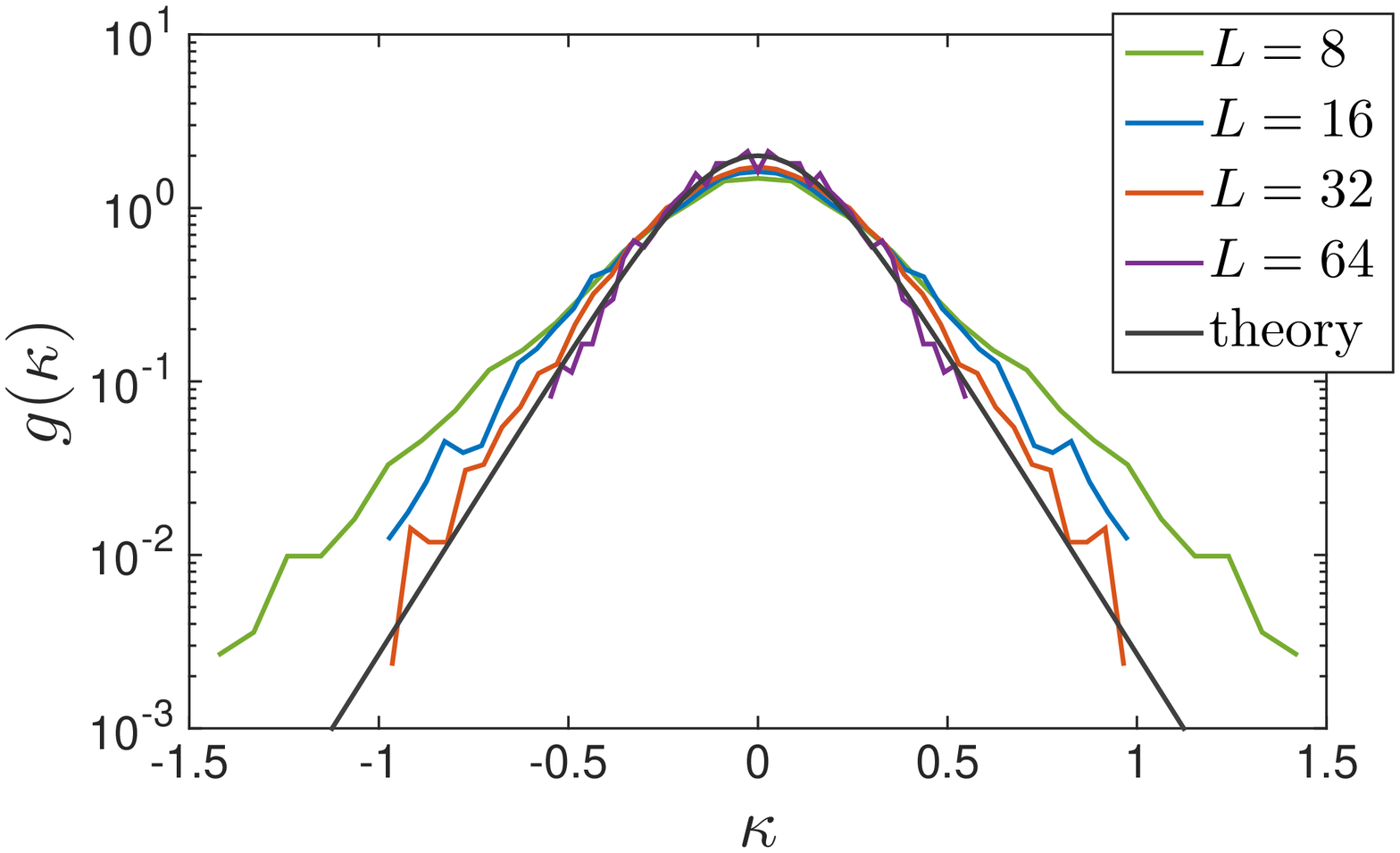}
\caption{\textcolor{black}{Distribution of the inverse decay length of floppy modes $g(\kappa)$ for free isostatic networks (top) and networks with pinned particles (bottom). The black solid line shows the theoretical prediction of Eq.(\ref{eq_pdflog}), based on free matrix multiplication. The fraction of pinned particles is $1/16$, constant for different system sizes. Inset: Distribution $\rho(\Lambda)$ of the singular value $\Lambda$. It converges to $1/\Lambda$ as implied by Eq.(\ref{eq_lambt}). }
\label{distr}}
\end{figure}

A quantitative characterization of displacement propagation is contained in the density of singular values $\rho(\Lambda)$. 
Left-right statistical symmetry implies that ${\cal T}_{RL}$  and its inverse have the same distribution, implying that $\rho(\Lambda)=\rho(1/\Lambda)/\Lambda^2$. Empirically, $\rho(\Lambda)$ is found to be very broad and appears to follow approximately $\rho(\Lambda)\sim 1/\Lambda$ in a wide region of $\Lambda$ values, as we report in Fig.~\ref{distr}. A more precise quantity are the inverse amplification/decay lengths $\kappa={\ln\Lambda}/{L}$, also called a Lyapounov exponent, whose distribution appears to converge  to a well defined limit $g(\kappa)$ in the thermodynamic limit $L \to \infty$ \cite{Sussman15}, as illustrated in Fig.~\ref{distr} (\textcolor{black}{top panel}). Our goal is to compute this distribution analytically. 

{\it The spectrum of slab matrices:} Imagine that the our network is sliced into $N_s$ thin slabs of thickness $\ell \sim 1$, such that $L=N_s \ell$. The $n^{th}$ slice is characterized by a transmission matrix ${\cal T}_{n,n-1}$. Translational invariance and left-right symmetry implies that the distribution of the singular values of each slab $\rho_s(\lambda)$ does not depend on $n$ and obeys $\rho_s(\lambda)=\rho_s(1/\lambda)/\lambda^2$. Because  each slice is a few particles wide, we expect that all $|\ln \lambda|$'s are of order unity independently of $L$. 

\begin{figure}[h!]
\centering
\includegraphics[width=.8\columnwidth]{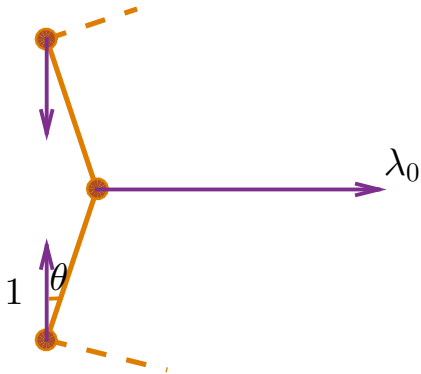}
\caption{A lever magnifying the motion by $\lambda_0$.
\label{lever}}
\end{figure}

We now argue that $\rho_s(\lambda)$ is in fact broadly distributed, due to the presence of local configurations that can greatly amplify the response by acting as a lever. An example of such a situation is shown in Fig.~\ref{lever}, for which imposing a unit vertical displacement on the left nodes leads to an horizontal displacement $\lambda_0=\tan\theta$ at the right node. In a random network, it is reasonable to assume that the angle $\theta$  is uniformly distributed in $[0,\frac{\pi}{2}]$. Thus the distribution $\rho_0(\lambda_0)$ of amplification factors follows as:
\be
\rho_0(\lambda_0)=\frac{2}{\pi}\frac{\rd\theta}{\rd\lambda_0}=\frac{2}{\pi}\frac{\rd\tan^{-1}\lambda_0}{\rd\lambda_0}=\frac{2}{\pi}\frac{1}{1+\lambda_0^2},
\label{eq_sig}
\ee
which is the Cauchy distribution. (Note that $\rho_0(\lambda_0)$ is indeed such that $\rho_0(\lambda_0)=\rho_0(1/\lambda_0)/\lambda_0^2$.)

\begin{figure}[h!]
\centering
\includegraphics[width=.8\columnwidth]{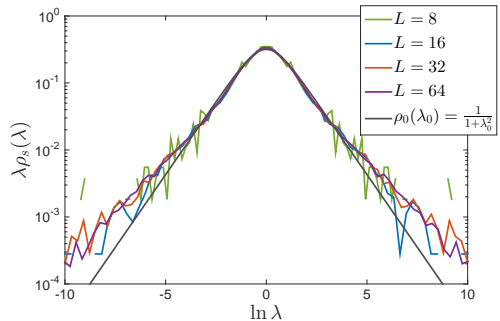}
\caption{Distribution of the singular values of layers of thickness $\ell_c=4$ for various size $L$ of the network, in log-log scale. The gray line shows the Cauchy distribution. 
\label{evslab}}
\end{figure}

From this argument, we expect $\rho_s(\lambda)$ to have fat tails, both for $\lambda \gg 1$ and $\lambda \ll 1$. We do indeed observe this behavior for all slab widths $\ell \sim O(1)$ that we probed (not shown). Moreover, we find empirically that for $\ell=\ell_c\approx 4$, $\rho_s(\lambda)$ is surprisingly well described by the above Cauchy distribution with no fitting parameters, i.e. $\rho_s(\lambda)\approx \rho_0(\lambda)$ for all values of $\lambda$, as demonstrated in Fig.~\ref{evslab}.  The value of $\ell_c$ presumably reflects the size of typical local lever, which may depend on the microscopic structure of the network. Henceforth we choose $\ell_c$ as our unit slab, such that $N_s=L/\ell_c$.

{\it Transmission matrix as a product of free matrices:} The transmission matrix can be exactly expressed a product of $N_s$ random slab matrices, as:
\be
{\cal T}_{LR}={\cal T}_{N,N-1}{\cal T}_{N-1,N-2}...{\cal T}_{10}.
\label{product}
\ee
Let us now make the strong assumption that those matrices are {\it mutually free} (i.e., in a nutshell, that the eigenbasis of the matrices ${\cal T}_{n,n-1}$ are 
independent for different $n$  \cite{Verdu}). Then, the singular value spectrum of the product can be expressed in terms of the singular value spectrum of each slab matrix \cite{Tucci11}, allowing one to express $\rho(\Lambda)$ in terms of $\rho_s(\lambda)$. A similar approximation is common in the context of one dimensional disordered conductors \cite{Beenakker97}. In the limit where the number $N_s$ of matrices in the product is large,  a non-trivial limit is reached for the probability density function $\textcolor{black}{p(t)}=P'(t)$ of 
the quantity $t \equiv \Lambda^{1/N}$.  This distribution can be computed from the so-called S-transform, defined via the following sequence of steps \cite{Tucci11}. First, one considers the distribution $\mu(\xi)$ of the real eigenvalues of the symmetric matrix ${\cal T}_{n-1,n}^\top{\cal T}_{n-1,n}$. This allows one to compute the following auxiliary function:
\be
\psi_\mu(z)=\int_0^\infty\frac{z \xi}{1-z\xi}\mu(\xi)\rd \xi.
\label{eq_psi}
\ee
Next, from the functional inverse $\chi_\mu$ of $\psi_\mu$, one defines the S-transform as
\be
S_\mu(z)=\frac{(1+z)\chi_{\mu}(z)}{z}; \qquad \chi_\mu(\psi_\mu(z))=z
\label{S}
\ee
Finally, the cumulative distribution $P(t)$ of the variable $t$ is obtained as the functional inverse of $P^{(-1)}(t)=\textcolor{black}{1/\sqrt{S_\mu(t-1)}}$.

We now apply this procedure to the present case. Since the eigenvalues $\xi$ of ${\cal T}_{n-1,n}^\top{\cal T}_{n-1,n}$ are the square of the Cauchy distributed singular values $\lambda$, we find:
\be
\mu(\xi)=\rho_0(\lambda)\frac{\rd\lambda}{\rd \xi}=\frac{1}{\pi}\frac{\xi^{-1/2}}{1+\xi}.
\label{eq_mu}
\ee
from which one easily obtains $\psi_\mu(z)=-1+1/(1+\sqrt{-z})$, $\chi_{\mu}(z)=-z^2/(1+z)^2$, $P=\frac{t^2}{1+t^2}$ and finally $\textcolor{black}{p(t)}=\frac{2t}{(1+t^2)^2}$. 
We define the inverse amplification/decay lengths $\tilde \kappa$ in slab units as $\tilde \kappa=\ell_c {\ln\Lambda}/{L} \equiv \ln t$. Hence we immediately deduce its probability density as:
\be
g({\tilde \kappa})=
\frac{2e^{2{\tilde \kappa}}}{(1+e^{2{\tilde \kappa}})^2}. 
\label{eq_pdflog}
\ee
This is the central result of our work. Note that $g(\tilde \kappa)$ is an even function in $\tilde \kappa$, as imposed by the left-right symmetry, and that 
it behaves as $e^{-2|\tilde \kappa|}$ for $|{\tilde \kappa}|\gg 1$. This prediction  is compared with numerical results in Fig.~\ref{distr}, and is found to 
be  good, especially considering that no fitting parameter has been introduced, once $\ell_c$ is chosen as above. Some discrepancies however appear, in particular an unpredicted cusp of the probability density seems to occur at $\tilde \kappa =0$, whereas our analytical prediction leads to a quadratic maximum (in the S.I. we show that the freeness assumption leads to a smooth maximum under a broad choice of slab spectrum $\rho_s(\Lambda)$). We discuss the origin of this discrepancy after  discussing the implications of our result.

{\it Inverse distribution of the singular value spectrum:} From the existence of a smooth thermodynamic limit for $\textcolor{black}{p(t)}$ or $g({\tilde \kappa})$, one can justify the 
$1/\Lambda$ behaviour for $\rho(\Lambda)$  illustrated in Fig.~\ref{distr}:
\be
\rho(\Lambda)=\textcolor{black}{p(t)}\rd t/\rd\Lambda=\frac{1}{N}\textcolor{black}{p}(\Lambda^{1/N}) \Lambda^{\frac{1}{N}-1} \xrightarrow{N \to \infty} \frac{\textcolor{black}{p}(1)}{\Lambda}.
\label{eq_lambt}
\ee

{\it Magnification of a random perturbation:} consider moving the nodes on the left edge of the system in a random direction, with a small magnitude $\delta  R_L $. What will be the characteristic amplification $\Lambda^*$ of the displacements at the right edge?
This response will be dominated by the maximally amplifying mode in the sample, whose associated $\kappa$-values is the largest. Since there are $M \sim L^{d-1}$
such eigenmodes, the largest one is set by  following an extreme value statistics argument:
\be
\int_{\kappa_{\max}}^\infty {\rm d}\kappa g(\kappa) \approx \frac{1}{L^{d-1}},
\ee
from which we deduce using Eq. \ref{eq_pdflog} that $\kappa_{\max}\approx \ln(M/2)/2\ell_c$, or else $\Lambda^*=e^{L \kappa_{max}} \sim L^{\frac{(d-1)L}{\ell_c}}$.
Thus, the magnification diverges extremely rapidly with the system size: on the right of the sample, the displacement will be of order $\delta  R_R\sim \delta R_L L^{\frac{(d-1)L}{\ell_c}}$. This is illustrated  in Fig.~\ref{magnif}(b), for which the displacement field is amplified by  $\approx 20,000$  for $L=16$ and $\ell_c=4$. \textcolor{black}{Note that the explosive modes associated to $\kappa > 0$ are excluded in the context of electronic transport, because of conservation laws that
are absent in the elastic problem considered here.}

{\it Penetration of floppy modes in the bulk:} Let us denote by $\{\delta {\bf R}_\alpha\}$ an orthonormal basis of the extended floppy modes extending from the left boundary, of dimension $M$.  A quantity characterizing how floppy modes penetrate in the bulk is \cite{Wyart05, Sussman15,Witten16} $f(x)=L \sum_{\alpha=1...M_{L}}  \langle \delta {\bf R}_\alpha(x)^2\rangle$ where $x$ is the distance from the left boundary, and $\langle \delta {\bf R}_\alpha(x)^2\rangle $ is the mean square displacement of  mode $\alpha$ at position $x$.  For homogeneously extended floppy modes, one expects that $f(x)$ tends to a constant for large $x$. Making the approximation  that the floppy modes associated with distinct singular values of ${\cal T}_{LR}$ -- such as those shown in Fig.~\ref{magnif} -- are orthonormal, we get for $x \ll L/2$ (such that the contribution of the decaying modes from the left boundary dominate):
\be
f(x)\propto L \int_{0}^{+\infty} {\rm d}\kappa g(\kappa) 2\kappa e^{-2\kappa x},
\ee
where the factor $2 \kappa$ comes from the $L_2$ normalisation of the modes in the $x$ direction. Changing variables from $\kappa \to q=x \kappa$, the
above integral yields: $f(x) \propto g(0) L/x^2$ for $1 \ll x \ll L/2$, indicating a slow, power-law decay of floppy modes inside the bulk. Our result explains the recent observations of \cite{Sussman15,Witten16} that $f(L/2)\sim 1/L$.

{\it Discussion:} Our freeness assumption predicts correctly the overall distribution of the Lyapounov exponents, but fails to capture its behavior near $\kappa=0$. This could be expected: although freeness is presumably a good approximation for short wavelength modes, it must fail for long wavelengths. In particular, purely translational modes must remain so after transmission by any slab transmission matrix ${\cal T}_{n,n-1}$, instead of becoming random as freeness would require. 
We thus expect that the accumulation of floppy modes near $\kappa=0$ stems from \textcolor{black}{translational modes and} long wavelength modes, which are less distorted by the  transmission process than we have assumed. We have indeed checked that near $\kappa=0$ floppy modes  strongly overlap with long wavelength modes, \textcolor{black}{as shown in S.I.}.

However, we can consider  practical situations where the transmitting material is clamped, for example by wrapping the system within a rigid wall parallel to the transmission direction (while removing other constraints to keep the system isostatic, see S.I. for an illustration). Such a constraint will make translational modes non-floppy, rendering the freeness approximation more accurate. We have tested this idea by studying transmission in a material where a fraction of the particles are pinned , essentially getting rid of the translational invariance responsible for the special role played by long wavelength modes. The procedure to pin particles is discussed in S.I. As illustrated in Fig.~\ref{distr}, our theoretical prediction indeed becomes accurate even for $\kappa\approx 0$ in such a set up.

{\it Conclusion:}
We have shown that the behavior of edge modes in disordered elastic networks is a transmission problem, which can be treated approximatively using recent mathematical results on the spectrum of the product of free random matrices. This approach rationalizes several previously unexplained observations. Moreover, it establishes that isostatic materials can generically act as {\it very effective levers}, which can considerably amplify motion without fine tuning the network geometry. It remains to be seen if such a property can be used to design artificial functional material, or if it is actually already used by Mother Nature, for example to confer to proteins their exceptional mechanical properties.


\begin{acknowledgements}
We thank C. Brito, E. DeGiuli, M. Pinson, R. Ravasio, T. Witten and S. Zamuner for discussions. M.W. thanks the Swiss National Science Foundation for support under Grant No. 200021-165509 and the Simons Foundation Grant ($\#$454953 Matthieu Wyart). \textcolor{black}{LY was supported in part by the National Science Foundation under Grant No. NSF PHY11-25915. This material is based upon work performed using computational resources supported by the ``Center for Scientific Computing at UCSB'' and NSF Grant CNS-0960316. }
\end{acknowledgements}

\bibliography{Wyartbibnew}

\renewcommand{\theequation}{S\arabic{equation}}
\setcounter{equation}{0}
\setcounter{figure}{0}
\renewcommand{\thefigure}{S\arabic{figure}}
\newpage
\section{Supplementary Information}

\subsection{Transmission matrices}

The coupling matrices ${\cal T}$ can be formally computed in the following way. We first calculate the $M_0$ extended floppy modes of the stiffness matrix, 
\be
{\cal M}|\phi^0_{\alpha}\rangle=0
\label{eq_floppy}
\ee
$\alpha=1,2,...,M_0$, and $|\phi^0_{\alpha}\rangle$ are vectors in $Nd$ displacement space. 

We then project these floppy modes onto the boundary nodes subspace $|\phi'_{\alpha}\rangle$, which are $(M+M_R)d$ dimensional. After orthonormalization, we obtain $M_0$ vectors $|\phi_{\alpha}\rangle$ in $(M+M_R)d$ dimensional subspace, 
which form a basis for any displacements of the boundary nodes along the floppy modes, 
\be
\left|\ba{c}\delta{\bf R}_L\\\delta{\bf R}_R\ea\right\rangle=\sum_{\alpha}c_{\alpha}|\phi_\alpha\rangle.
\label{eq_decompose}
\ee
For a given perturbing displacement on the left boundary $|\delta{\bf R}_L\rangle$, the projection onto different floppy modes $c_{\alpha}$ is determined as $Md\geq M_0$, 
\be
c_\alpha=\left\langle\phi_{\alpha}\middle|\ba{c}\delta{\bf R}_L\\\delta{\bf R}_R\ea\right\rangle=\left\langle\phi_{\alpha}\middle|\ba{c}\delta{\bf R}_L\\{\bf 0}\ea\right\rangle+\left\langle\phi_{\alpha}\middle|\ba{c}{\bf 0}\\\delta{\bf R}_R\ea\right\rangle
\label{eq_coeff}
\ee
With the help of the projection operators ${\mathbbm 1}_R$, 
\be
{\mathbbm 1}_R\left|\ba{c}\delta{\bf R}_L\\\delta{\bf R}_R\ea\right\rangle=\left|\ba{c}{\bf 0}\\\delta{\bf R}_R\ea\right\rangle
\label{eq_pr}
\ee
and similarly ${\mathbbm 1}_L$, we can derive a linear relation between the response field on the right boundary $|\delta{\bf R}_R\rangle$ and $|\delta{\bf R}_L\rangle$,
\begin{widetext}
\be
\left|\ba{c}{\bf 0}\\\delta{\bf R}_R\ea\right\rangle=\left(\left[{\mathbbm 1}_R-\sum_\alpha{\mathbbm 1}_R|\phi_\alpha\rangle\langle\phi_\alpha|\right]^{-1}{\mathbbm 1}_R\sum_{\alpha}|\phi_\alpha\rangle\langle\phi_\alpha|\right)\left|\ba{c}\delta{\bf R}_L\\{\bf 0}\ea\right\rangle
\label{eq_sol}
\ee
\end{widetext}
where the inverse is a pseudo-inverse, only in the subspace of the matrix. We further define projection operators, 
\be
{\cal P}\equiv\sum_{\alpha}|\phi_\alpha\rangle\langle\phi_\alpha|, 
\label{eq_p}
\ee
onto the subspace of the floppy modes, and 
\be
{\cal Q}\equiv{\mathbbm 1}-{\cal P},
\label{eq_q}
\ee
the complimentary space of the floppy modes on the boundary nodes. 
The linear coupling matrices between the left and right boundary displacements can be formulated as,
\be
{\cal \widetilde T}_{RL}=({\mathbbm 1}_R{\cal Q}{\mathbbm 1}_R)^{-1}{\mathbbm 1}_R{\cal P}{\mathbbm 1}_L;
\label{eq_trl}
\ee
\be
{\cal \widetilde T}_{LR}=({\mathbbm 1}_L{\cal Q}{\mathbbm 1}_L)^{-1}{\mathbbm 1}_L{\cal P}{\mathbbm 1}_R.
\label{eq_tlr}
\ee
Notice that ${\mathbbm 1}_L$ and ${\mathbbm 1}_R$ are also complementary projection operators, ${\mathbbm 1}={\mathbbm 1}_L+{\mathbbm 1}_R$. 
Any localized dangling mode $|\phi_\alpha\rangle$ does not contribute to the coupling matrix ${\cal \widetilde T}$: if in the left boundary nodes, ${\mathbbm 1}_R|\phi_\alpha\rangle=0$; while in the right boundary  nodes, $\langle\phi_\alpha|{\mathbbm 1}_L=0$. 
${\cal \widetilde T}_{LR}{\cal \widetilde T}_{RL}$ is therefore a projection operator onto the subspace of the floppy modes disregarding the dangling ones, essentially due to a left-right boundary symmetry. By applying twice the linear transform Eq.(\ref{eq_lin}) to Eq.(\ref{eq_coeff}),
\be
{\cal \widetilde T}_{LR}{\cal \widetilde T}_{RL}{\mathbbm 1}_L|\phi_{\alpha}\rangle={\mathbbm 1}_L|\phi_{\alpha}\rangle
\label{eq_id}
\ee
for $\phi_{\alpha}$ in the space of extended modes. Equivalently, ${\cal \widetilde T}_{RL}{\cal \widetilde T}_{LR}$ is a projector on the same subspace of the floppy modes but presented in right boundary space. 

 In our main text, the notation ${\cal T}_{RL}$ is used for the restriction of ${\cal \widetilde T}_{RL}$ to the space orthogonal to its kernel (i.e. the space of extended modes).
 
\begin{figure}[h!]
\centering
\includegraphics[width=.8\columnwidth]{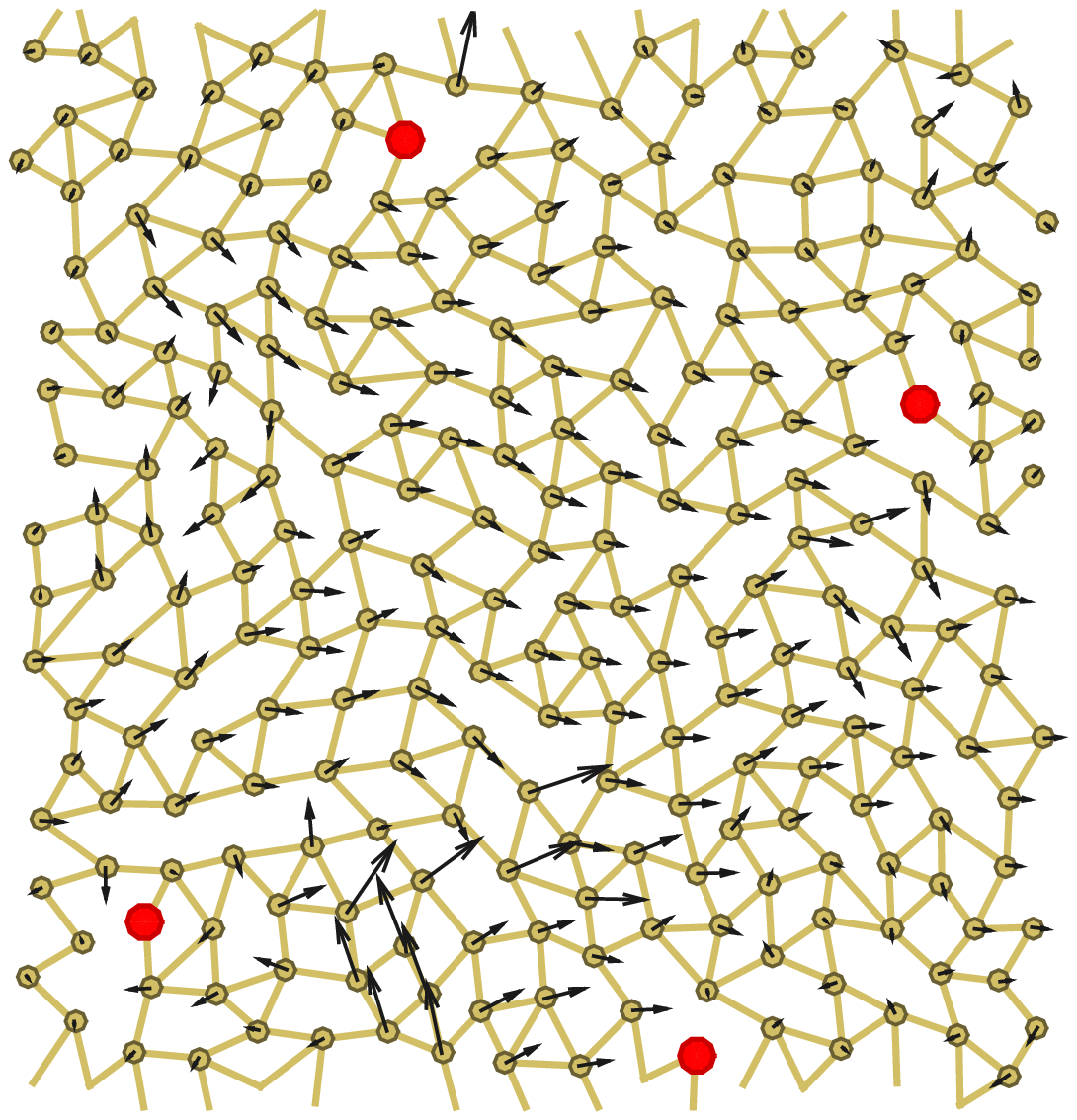}
\caption{
\textcolor{black}{An isostatic 2-dimensional network $z=4$ with four pinned particles, shown in red. The boundaries are open on the left and right hand sides of the network, and periodic in the vertical direction. Black arrows show a typical floppy mode penetrating the network.}
\label{pinned}}
\end{figure}

\textcolor{black}{
\subsection{Networks with pinned particles}
To eliminate floppy translational modes, we consider networks with pinned particles. 
We choose these particles randomly, starting from an isostatic network~\cite{Yan16}. To ensure the  the spatial homogeneity of constraints, we divide the network into small rectangles with one to two pinned particles chosen in each. Each time a particle is pinned, we remove $d$ bonds connecting it to its best connected neighbors, so that the resulted network is still isostatic. An example of such a network is shown in Fig.~\ref{pinned}. 
}

\textcolor{black}{
Following the formalism described in the previous section, we can compute the singular values for the pinned network. Notice that the structure matrix $\cal{S}$ of the pinned network includes bonds connecting the pinned and free particles. The resulting distributions are shown in the bottom panel of Fig.~\ref{distr} in the main text. The cusp at $\kappa=0$ is seems to disappear, and the distribution nicely converges  to our theoretical prediction.
}


\subsection{A general perturbative expansion}

In this section we specialize the general result of \cite{Tucci11} to the case where the eigenvalues $\xi$ are all close to unity. More precisely, we 
write $\xi : = e^{\sqrt{\epsilon} \zeta}$ with  $\epsilon \to 0$, and where $\zeta$ has an arbitrary even distribution $P(\zeta)=P(-\zeta)$. We will denote its 
variance as $\zeta_2$ and its fourth moment as $\zeta_4$. 

The first task is to compute $\psi(z)$ as defined in the text, Eq. (\ref{eq_psi}), up to order $\epsilon^2$. One readily finds: 
\be 
\psi(z) = \frac{z}{1-z} + \left[\epsilon \frac{\zeta_2}{2} + \epsilon^2 \frac{\zeta_4}{24}\right] \frac{z(1+z)}{(1-z)^3} + 
\epsilon^2 \zeta_4 \frac{z^2(1+z)}{(1-z)^5} + O(\epsilon^4). 
\ee 
Now, one should invert this relation to find $\chi$ such that $\chi(\psi(z))=z$. Proceeding order by order in $\epsilon$ leads to: 
\bea 
\chi(z)= \frac{z}{1+z} \big( 1 - \left[\epsilon \frac{\zeta_2}{2} + \epsilon^2 \frac{\zeta_4}{24}\right] (1+ 2z) + \nonumber\\ 
\epsilon^2 \frac{\zeta_2^2}{4} (1+z)(1+2z) \left[(1+4z) - 2 \varkappa z\right] \big), 
\eea
where we introduced the ratio $\varkappa = \zeta_4/\zeta_2^2$. From $\chi$ one deduces the S-transform as $S(z) = (1+z)\chi(z)/z$ and finally 
$F(z)=1/\sqrt{S(z-1)}$ to order $\epsilon^2$: 
\bea 
&&F(z) = 1 + \left[\epsilon \frac{\zeta_2}{2} + \epsilon^2 \frac{\zeta_4}{24}\right](z-\frac12)\nonumber \\
&&+ \epsilon^2 \frac{\zeta_2^2}{4} (2z-1) \left[\varkappa z(z-1) - \frac{16 z^2 - 18 z + 3}{8}\right] 
\eea 
This allows us to determine the edges of the distribution as $t_{\text{min}}=F(0)$ and $t_{\text{max}}=F(1)$, which are found to be at a distance $O(\epsilon)$ of $t=1$.   
The cumulative distribution $P(t)$ of the eigenvalues of the asymptotic product as the inverse of $F(z)$: 
\be 
F(P(t))=t. 
\ee 
Taking the derivative of this expression with  respect to $t$ and introducing the pdf $f(t)=P'(t)$ one gets: 
\be 
f(t) \left.\frac{\partial F}{\partial z}\right|_{z = P(t)} = 1 
\ee 
From the above expression of $F(z)$, one derives, setting $z := (1+u)/2$ 
\be 
\frac{\partial F}{\partial z} = \left[\epsilon \frac{\zeta_2}{2} + \epsilon^2 \frac{\zeta_4}{24}\right] + \epsilon^2 \frac{\zeta_2^2}{8} 
\left[ \varkappa (3 u^2 - 1) - (6u^2 -u +1) \right]. 
\ee 
To order $\epsilon^2$, one therefore only needs the relation between $t$ and $u$ at order $\epsilon$, which reads: 
\be 
t = 1 + \frac14 \epsilon \zeta_2  u. 
\ee 
It turns out to be more convenient to work with the variable $x$ such that $\epsilon \zeta_2 x := 4 \ln t$. In particular, 
due to the symmetry $t \to 1/t$, one must find that the distribution of $x$ is even. The final result is, to leading order: 
\be 
f(x) = 2 + \epsilon\frac{\zeta_2}{2} \left[(\varkappa - 1) + \frac{\zeta_2}{4} (6 - 3 \varkappa) x^2\right], 
\ee 
for $x \in [-1+\epsilon\zeta_2/4,1-\epsilon\zeta_2/4]$, and zero otherwise. By inspection, this distribution is indeed even in $x$. 
\\
The conclusion of this computation is that: 
\begin{itemize} 
\item To lowest order in $\epsilon$, the distribution of eigenvalues $t$ is universal and uniform around unity, in a symmetric interval of size $\epsilon \zeta_2/2$. This 
ties up with the result obtained in the context of 1-d disordered conductors in the corresponding limit, see \cite{Beenakker97}, section III.B. 
\item To next order, we find that the distribution acquires a parabolic shape, with a curvature that depends on $\varkappa$, i.e. the kurtosis of the distribution 
of $\zeta \propto \ln \xi$. For a bimodal distribution, one finds $\varkappa=1$ and a positive curvature, such that $t=1$ is a minimum of the distribution. 
As soon as $\varkappa > 2$ (for example for a Gaussian distribution of $\zeta$), the curvature is negative and $t=1$ is the mode of the distribution. The case of a 
Cauchy distribution for $\lambda=\sqrt{\xi}$, as in the text, corresponds to $\varkappa = 5$, such that the distribution indeed is expected to be bell shaped in this case. 
\item One also concludes that  if freeness holds, the only way to obtain a cusp in the distribution at $t=1$ is to have an infinite kurtosis $\varkappa$ at the microscopic level. This 
is clearly not the case of our ``slab'' matrices, see Fig.~ \ref{evslab}. 
\end{itemize} 

\begin{figure}[h!]
\centering
\includegraphics[width=.8\columnwidth]{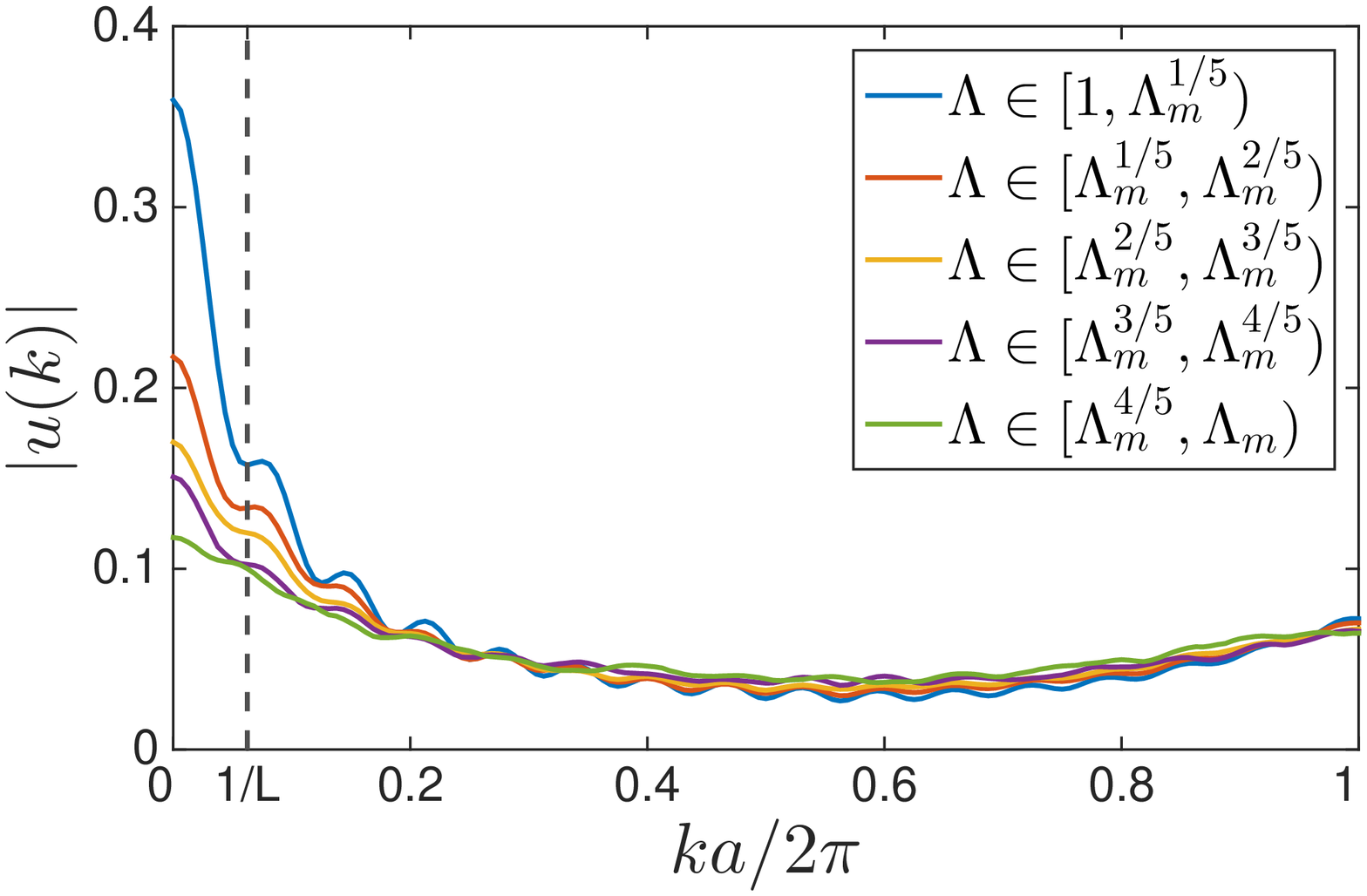}\\
\includegraphics[width=.8\columnwidth]{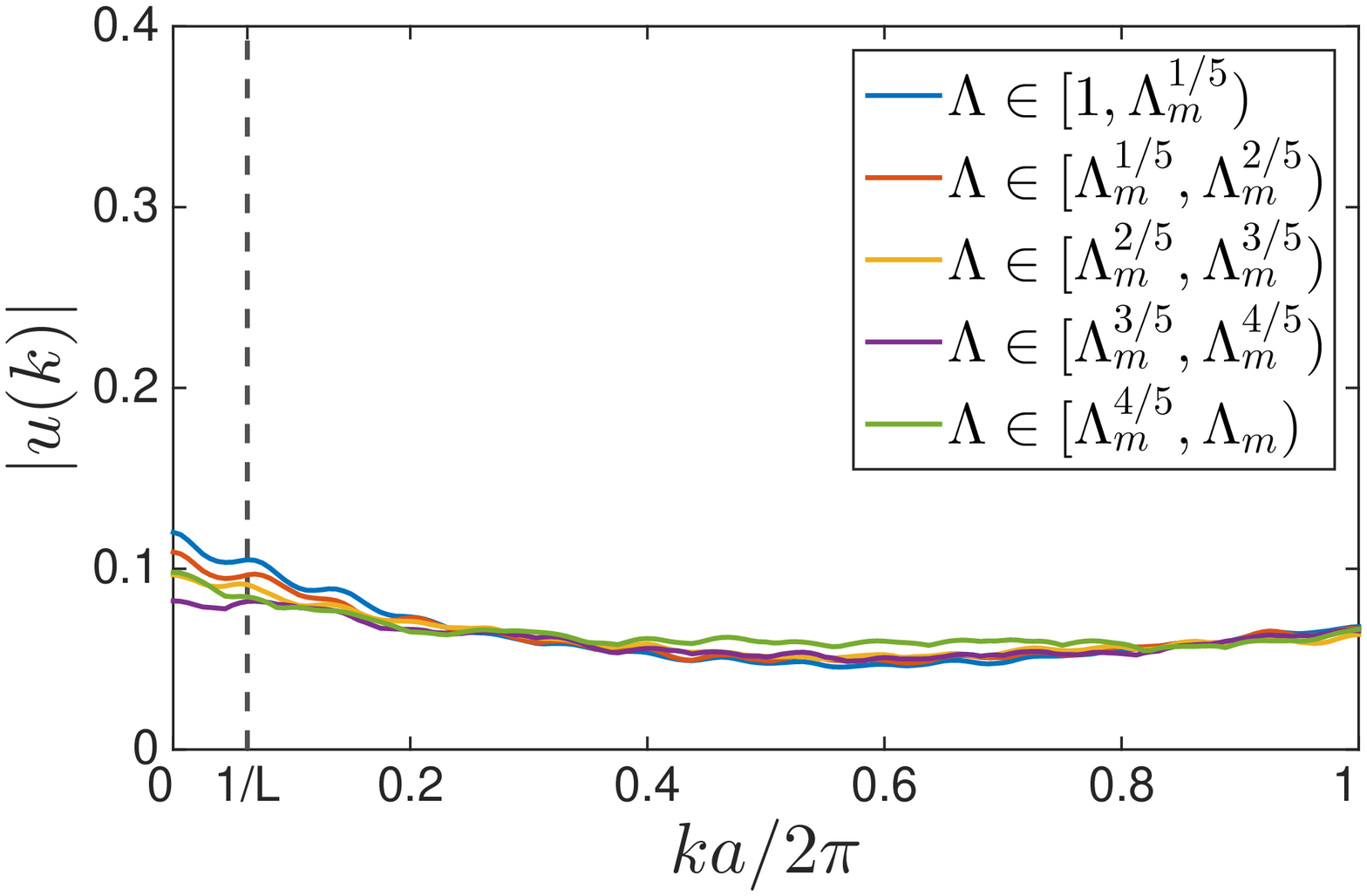}\\
\caption{
\textcolor{black}{Fourier transform of the displacement fields $u(k)$ (defined in the text) corresponding to different singular value range $\Lambda$ for free isostatic networks (top) and networks with pinned particles (bottom). The fraction of pinned particles is $0.12$. $L=16$,  and $a$ is the typical distance between the particles.}
\label{fourier}}
\end{figure}

\textcolor{black}{
\subsection{Overlap with sinusoidal modes}
To check that the coupling between translational (or long wavelength) modes and floppy modes is responsible for the apparent cusp in the distribution $g(\kappa)$, we measure the Fourier components of the floppy modes near $\Lambda=1$. To do so, we use the following recipe. Let us consider that the $x$-axis is in the horizontal direction which connects the two open boundaries, and that the $y$-axis is the vertical direction which is periodic. 
1. Noticing that the amplitude of the floppy modes varies exponentially along the $x$ direction, we first rescale the field by $\Lambda^{x/L}$, so that the displacements are now of the similar magnitude when $x$ varies, so that each slab  contributes similarly  to the Fourier transform. 
2. We normalize the mode, and get $\delta\vec{r} = \delta r_x \hat{x} + \delta r_y \hat{y}$ at each node $j$. 
3. We make the Fourier transformation along the direction parallel to the open boundaries by $\vec{u}(k) = \sum_j \delta r_x(j)e^{-iky_j} \hat{x} + \delta r_y(j)e^{-iky_j} \hat{y}$. 
4. Finally, we compute the absolute value of $\vec{u}(k)$. 
}

\textcolor{black}{
The results averaged over various $\Lambda$ are shown in Fig.~\ref{fourier}. In the original  networks (no pinned particles), for modes near $\Lambda=1$ we clearly see a strong peak at $k=0$ and $k=2\pi n/La$ for small integers $n$. The closer to $\Lambda=1$, the stronger is the effect. When a fraction of the particles are pinned, this strong coupling with translational modes and sinusoidal modes disappear. The  Fourier decomposition is then close to flat, as it should be in the free matrix approximation where eigenvectors are random.
}

\end{document}